\documentclass[a4paper,11pt]{article}
\usepackage[T1]{fontenc} 

\makeatletter
\gdef\@fpheader{ }
\gdef\@journal{ }
\makeatother
\RequirePackage{amsmath}
\RequirePackage{amssymb}
\RequirePackage{epsfig}
\RequirePackage{graphicx}
\RequirePackage[numbers,sort&compress]{natbib}
\RequirePackage{color}
\RequirePackage[colorlinks=true
,urlcolor=blue
,anchorcolor=blue
,citecolor=blue
,filecolor=blue
,linkcolor=blue
,menucolor=blue
,pagecolor=blue
,linktocpage=true
,pdfproducer=medialab
,pdfa=true
]{hyperref}

\newif\ifnotoc\notocfalse
\newif\ifemailadd\emailaddfalse
\newif\iftoccontinuous\toccontinuousfalse
\makeatletter
\def\@subheader{\@empty}
\def\@keywords{\@empty}
\def\@abstract{\@empty}
\def\@xtum{\@empty}
\def\@dedicated{\@empty}
\def\@arxivnumber{\@empty}
\def\@collaboration{\@empty}
\def\@collaborationImg{\@empty}
\def\@proceeding{\@empty}
\def\@preprint{\@empty}

\newcommand{\subheader}[1]{\gdef\@subheader{#1}}
\newcommand{\keywords}[1]{\if!\@keywords!\gdef\@keywords{#1}\else%
\PackageWarningNoLine{\jname}{Keywords already defined.\MessageBreak Ignoring last definition.}\fi}
\renewcommand{\abstract}[1]{\gdef\@abstract{#1}}
\newcommand{\dedicated}[1]{\gdef\@dedicated{#1}}
\newcommand{\arxivnumber}[1]{\gdef\@arxivnumber{#1}}
\newcommand{\proceeding}[1]{\gdef\@proceeding{#1}}
\newcommand{\xtumfont}[1]{\textsc{#1}}
\newcommand{\correctionref}[3]{\gdef\@xtum{\xtumfont{#1} \href{#2}{#3}}}
\newcommand\jname{JHEP}
\newcommand\acknowledgments{\section*{Acknowledgments}}

\newcommand\preprint[1]{\gdef\@preprint{\hfill #1}}

\makeatother

\newcommand\note[2][]{%
\if!#1!%
\stepcounter{footnote}\footnotetext{#2}%
\else%
{\renewcommand\thefootnote{#1}%
\footnotetext{#2}}%
\fi}

\makeatletter
\newtoks\auth@toks
\renewcommand{\author}[2][]{%
  \if!#1!%
    \auth@toks=\expandafter{\the\auth@toks#2\ }%
  \else
    \auth@toks=\expandafter{\the\auth@toks#2$^{#1}$\ }%
  \fi
}
\makeatother
\makeatletter
\newtoks\affil@toks\newif\ifaffil\affilfalse
\newcommand{\affiliation}[2][]{%
\affiltrue
  \if!#1!%
    \affil@toks=\expandafter{\the\affil@toks{\item[]#2}}%
  \else
    \affil@toks=\expandafter{\the\affil@toks{\item[$^{#1}$]#2}}%
  \fi
}
\makeatother
\makeatletter
\newtoks\email@toks\newcounter{email@counter}%
\newcommand{\emailAdd}[1]{%
\emailaddtrue%
\ifnum\theemail@counter>0\email@toks=\expandafter{\the\email@toks, \@email{#1}}%
\else\email@toks=\expandafter{\the\email@toks\@email{#1}}%
\fi\stepcounter{email@counter}}
\newcommand{\@email}[1]{\href{mailto:#1}{\tt #1}}
\makeatother

\makeatletter
\newcommand*\collaboration[1]{\gdef\@collaboration{#1}}
\newcommand*\collaborationImg[2][]{\gdef\@collaborationImg{#2}}
\makeatletter
\newcommand\afterLogoSpace{\smallskip}
\newcommand\afterSubheaderSpace{\vskip3pt plus 2pt minus 1pt}
\newcommand\afterProceedingsSpace{\vskip21pt plus0.4fil minus15pt}
\newcommand\afterTitleSpace{\vskip23pt plus0.06fil minus13pt}
\newcommand\afterRuleSpace{\vskip23pt plus0.06fil minus13pt}
\newcommand\afterCollaborationSpace{\vskip3pt plus 2pt minus 1pt}
\newcommand\afterCollaborationImgSpace{\vskip3pt plus 2pt minus 1pt}
\newcommand\afterAuthorSpace{\vskip5pt plus4pt minus4pt}
\newcommand\afterAffiliationSpace{\vskip3pt plus3pt}
\newcommand\afterEmailSpace{\vskip16pt plus9pt minus10pt\filbreak}
\newcommand\afterXtumSpace{\par\bigskip}
\newcommand\afterAbstractSpace{\vskip16pt plus9pt minus13pt}
\newcommand\afterKeywordsSpace{\vskip16pt plus9pt minus13pt}
\newcommand\afterArxivSpace{\vskip3pt plus0.01fil minus10pt}
\newcommand\afterDedicatedSpace{\vskip0pt plus0.01fil}
\newcommand\afterTocSpace{\bigskip\medskip}
\newcommand\afterTocRuleSpace{\bigskip\bigskip}
\newlength{\affiliationsSep}
\newcommand\beforetochook{\pagestyle{myplain}\pagenumbering{roman}}

\DeclareFixedFont\trfont{OT1}{phv}{b}{sc}{11}

\renewcommand\maketitle{
\pagestyle{empty}
\thispagestyle{titlepage}
\setcounter{page}{0}
\noindent{\small\scshape\@fpheader}\@preprint\par

\afterLogoSpace
\if!\@subheader!\else\noindent{\trfont{\@subheader}}\fi
\afterSubheaderSpace
\if!\@proceeding!\else\noindent{\sc\@proceeding}\fi
\afterProceedingsSpace
{\LARGE\flushleft\sffamily\bfseries\@title\par}
\afterTitleSpace
\hrule height 1.5\p@%
\afterRuleSpace
\if!\@collaboration!\else
{\Large\bfseries\sffamily\raggedright\@collaboration}\par
\afterCollaborationSpace
\fi
\if!\@collaborationImg!\else
{\normalsize\bfseries\sffamily\raggedright\@collaborationImg}\par
\afterCollaborationImgSpace
\fi
{\bfseries\raggedright\sffamily\the\auth@toks\par}
\afterAuthorSpace
\ifaffil\begin{list}{}{%
\setlength{\leftmargin}{0.28cm}%
\setlength{\labelsep}{0pt}%
\setlength{\itemsep}{\affiliationsSep}%
\setlength{\topsep}{-\parskip}}
\itshape\small%
\the\affil@toks
\end{list}\fi
\afterAffiliationSpace
\ifemailadd 
\noindent\hspace{0.28cm}\begin{minipage}[l]{.9\textwidth}
\begin{flushleft}
\textit{E-mail:} \the\email@toks
\end{flushleft}
\end{minipage}
\else 
\PackageWarningNoLine{\jname}{E-mails are missing.\MessageBreak Plese use \protect\emailAdd\space macro to provide e-mails.}
\fi
\afterEmailSpace
\if!\@xtum!\else\noindent{\@xtum}\afterXtumSpace\fi
\if!\@abstract!\else\noindent{\renewcommand\baselinestretch{.9}\textsc{Abstract:}}\ \@abstract\afterAbstractSpace\fi
\if!\@keywords!\else\noindent{\textsc{Keywords:}} \@keywords\afterKeywordsSpace\fi
\if!\@arxivnumber!\else\noindent{\textsc{ArXiv ePrint:}} \href{http://arxiv.org/abs/\@arxivnumber}{\@arxivnumber}\afterArxivSpace\fi
\if!\@dedicated!\else\vbox{\small\it\raggedleft\@dedicated}\afterDedicatedSpace\fi
\ifnotoc\else
\iftoccontinuous\else\newpage\fi
\beforetochook\hrule
\tableofcontents
\afterTocSpace
\hrule
\afterTocRuleSpace
\fi
\setcounter{footnote}{0}
\pagestyle{myplain}\pagenumbering{arabic}
} 

\renewcommand{\baselinestretch}{1.1}\normalsize
\divide\@tempdima\baselineskip
\@tempcnta=\@tempdima
\skip\@mpfootins = \skip\footins
\renewcommand{\@dotsep}{10000}

\newcommand\ps@myplain{
\pagenumbering{arabic}
\renewcommand\@oddfoot{\hfill-- \thepage\ --\hfill}
\renewcommand\@oddhead{}}
\let\ps@plain=\ps@myplain

\newcommand\ps@titlepage{\renewcommand\@oddfoot{}\renewcommand\@oddhead{}}

\numberwithin{equation}{section}

\renewcommand\section{\@startsection{section}{1}{\z@}%
                                   {-3.5ex \@plus -1.3ex \@minus -.7ex}%
                                   {2.3ex \@plus.4ex \@minus .4ex}%
                                   {\normalfont\large\bfseries}}
\renewcommand\subsection{\@startsection{subsection}{2}{\z@}%
                                   {-2.3ex\@plus -1ex \@minus -.5ex}%
                                   {1.2ex \@plus .3ex \@minus .3ex}%
                                   {\normalfont\normalsize\bfseries}}
\renewcommand\subsubsection{\@startsection{subsubsection}{3}{\z@}%
                                   {-2.3ex\@plus -1ex \@minus -.5ex}%
                                   {1ex \@plus .2ex \@minus .2ex}%
                                   {\normalfont\normalsize\bfseries}}
\renewcommand\paragraph{\@startsection{paragraph}{4}{\z@}%
                                   {1.75ex \@plus1ex \@minus.2ex}%
                                   {-1em}%
                                   {\normalfont\normalsize\bfseries}}
\renewcommand\subparagraph{\@startsection{subparagraph}{5}{\parindent}%
                                   {1.75ex \@plus1ex \@minus .2ex}%
                                   {-1em}%
                                   {\normalfont\normalsize\bfseries}}

\def\fnum@figure{\textbf{\figurename\nobreakspace\thefigure}}
\def\fnum@table{\textbf{\tablename\nobreakspace\thetable}}

\long\def\@makecaption#1#2{%
  \vskip\abovecaptionskip
  \sbox\@tempboxa{\small #1. #2}%
  \ifdim \wd\@tempboxa >\hsize
    \small #1. #2\par
  \else
    \global \@minipagefalse
    \hb@xt@\hsize{\hfil\box\@tempboxa\hfil}%
  \fi
  \vskip\belowcaptionskip}

\renewenvironment{thebibliography}[1]{%
\begin{oldthebibliography}{#1}%
\small%
\raggedright%
\setlength{\itemsep}{5pt plus 0.2ex minus 0.05ex}%
}%
{%
\end{oldthebibliography}%
}

\begin{document}



\renewcommand{\thefootnote}{\fnsymbol{footnote}}

\title{\boldmath A $1+5$-dimensional gravitational-wave solution: curvature singularity and
spacetime singularity}%


\author[a]{Yu-Zhu Chen,}
\author[b,a]{Wen-Du Li,}
\author[a*]{and Wu-Sheng Dai}\note{daiwusheng@tju.edu.cn.}


\affiliation[a]{Department of Physics, Tianjin University, Tianjin 300350, P.R. China}
\affiliation[b]{Theoretical Physics Division, Chern Institute of Mathematics, Nankai University, Tianjin, 300071, P. R. China}









\abstract{We solve a $1+5$-dimensional cylindrical gravitational-wave solution of the
Einstein equation, in which there are two curvature singularities. Then we
show that one of the curvature singularities can be removed by an extension of
the spacetime. The result exemplifies that the curvature singularity is not
always a spacetime singularity; in other words, the curvature singularity
cannot serve as a criterion for spacetime singularities.
}

\maketitle
\flushbottom


\section{Introduction}

We find a $1+5$-dimensional cylindrical vacuum\ gravitational-wave solution of
the Einstein equation. Such a spacetime has both the curvature singularity,
the singularity of the Riemann tensor components $R_{\text{ }ikl}^{j}$, and
the scalar polynomial curvature singularity, the singularity of the scalars
formed out of the curvatures such as $R$, $R^{ij}R_{ij}$, and $R_{\text{ }%
ikl}^{j}R_{j}^{\text{ }ikl}$. In this $1+5$-dimensional spacetime, there is
a\ singularity hypersurface formed by singular points behaving like a horizon.

The spacetime singularity is an important problem in gravity theory.
Nevertheless, even the definition of the spacetime singularity is still an
open question. Many attempts have been made to seek a general criterion for
the spacetime singularity\textbf{.}

There is no general criterion to judge whether a spacetime is singularity-free
\cite{wald2010general}. By intuition, the singularity should involve infinite
curvatures. However, Geroch constructs some singular spacetimes whose Riemann
curvatures are bounded everywhere; that is to say, the singularity cannot be
determined by the Riemann curvature \cite{geroch1968singularity}. Some efforts
are made to define a singular boundary which is coordinate independent, such
as the "g-boundary" \cite{geroch1968local} and the "b-boundary"
\cite{schmidt1971new}. These prescriptions and other similar procedures,
however, produce pathological topological boundaries in simple examples
\cite{johnson1977bundle,geroch1982singular,wald2010general}. Hence, the
singular boundary cannot be generally defined. Up to now, a generally accepted
definition of the spacetime singularity is based on timelike- or null-geodesic
incompleteness \cite{hawking1974large}. Based on the geodesic incompleteness,
Hawking and Penrose prove a series of singularity theorems
\cite{hawking1974large}. Nevertheless, Geroch also constructs a singular
spacetime whose geodesics are all complete, so that these singularities cannot
be judged by the geodesic completeness \cite{geroch1968singularity}. In short,
the reason why it is hard to present a general criterion for the singularity
is that the spacetime may have an infinite variety of possible pathological
behaviors \cite{wald2010general}.

On the contrary, one can easily judge that a spacetime has a singularity if
the spacetime has one kind of pathological behaviors. It is a common
impression that if the curvature is singular at some point, the spacetime is
singular. It should be emphasized that an orthonormal frame is necessary. In
an orthonormal frame, the metric of a $1+n$-dimensional spacetime reads
$ds^{2}=-\left(  \theta^{0}\right)  ^{2}+\sum_{i=1}^{n}\left(  \theta
^{i}\right)  ^{2}$ with $\theta^{0}$ and $\theta^{i}$ the $1$-form on the
manifold, which are independent of the choice of coordinates. As a result, the
curvature in an orthonormal frame is independent of the choice of the
coordinates \cite{lam2000lectures,misner1973gravitation}.

In this paper, we exemplify by the $1+5$-dimensional gravitational-wave
solution obtained in the present paper that a curvature singularity is not
always a spacetime singularity. Concretely, we show that the curvature
singularity can be removed by an extension of the solution to the maximal
manifold. This result tells us that the curvature singularity is not always a
spacetime singularity and thus cannot serve as a criterion for spacetime singularities.

We show that this $1+5$-dimensional cylindrical vacuum\ solution describes a
gravitational wave, through calculating energy-momentum pseudotensors
\cite{landau2013classical}. The gravitational wave is always an important
problem in gravity theory \cite{sathyaprakash2009physics}. The existence of
the gravitational wave is confirmed experimentally
\cite{abbott2016observation,aasi2014gravitational}. As the key source of the
gravitational wave, the black-hole binaries are allowed stable, robust
simulations of the merger process\textbf{ }\cite{centrella2010black} based on
the advance in numerical relativity \cite{centrella2010black}. New sources of
the gravitational waves are discussed \cite{senatore2014new}. The application
of the gravitational wave as a probe with the first order electroweak phase
transition is also considered \cite{kakizaki2015gravitational}.

Moreover, we also give a brief discussion on the $1+n$-dimensional case.

Higher dimensional gravity is an important issue in the theory of gravity.
There are also some studies on the cylindrical spacetime, such as a
cylindrical solution in the mimetic gravity theory
\cite{momeni2016cylindrical} and higher dimensional cylindrical or Kasner type
electrovacuum solutions \cite{delice2013higher}. For higher dimensional
spacetime, there are some exact solutions of the Einstein equation, such as a
rotating black ring solution in five dimensions \cite{emparan2002rotating},
general Kerr-NUT-AdS metrics in arbitrary dimensions \cite{chen2006general}%
,\ and the exact solutions of Einstein-Maxwell gravity
\cite{rahaman2006thin,aliev2006rotating}. In Refs.
\cite{hollands2007higher,vacaru2010general}, the property of the exact
solution of the Einstein equation in higher dimensions are discussed. In Refs.
\cite{tomizawa2006vacuum1,tomizawa2006vacuum2}, an inverse scattering method
for solving the vacuum solution of five dimensional Einstein equation are
provided. In the brane world model, the four-dimensional world is regarded as
a brane embedded in higher-dimensional spacetime
\cite{padmanabhan2010gravitation,mukohyama2000brane,gherghetta2000living}.
Beyond general relativity, there are also some modified gravity theories are
considered, such as the Lanczos--Lovelock model \cite{padmanabhan2013lanczos},
and the Einstein-Gauss-Bonnet theory \cite{dotti2007static}.

In section \ref{solution}, a $1+5$-dimensional cylindrical vacuum\ solution of
the Einstein equation is presented. In section \ref{singularity}, the
singularity of this $1+5$-dimensional cylindrical\textbf{ }spacetime\ is
analyzed. In section \ref{extension}, we make an extension of this
$1+5$-dimensional cylindrical spacetime to remove the curvature singularity.
In section \ref{hypersurface}, we show that in this $1+5$-dimensional
cylindrical spacetime there exists a horizon-like singularity hypersurface. In
section \ref{gravitational wave}, we show that the solution presented in this
paper describes a gravitational wave. In section \ref{1pn}, we give a brief
discussion on $1+n$-dimensional cylindrical vacuum\ solutions. The conclusions
are summarized in section \ref{Conclusion}.

\section{$1+5$-dimensional cylindrical vacuum\ solution \label{solution}}

In this section, we find a $1+5$-dimensional cylindrical vacuum\ solution of
the Einstein equation.

Choose the metric of a $1+5$-dimensional cylindrical spacetime as
\begin{equation}
ds^{2}=e^{2\Phi\left(  t,\rho\right)  }\left(  -dt^{2}+d\rho^{2}\right)
+e^{-2\psi\left(  t,\rho\right)  }\rho^{2}d\Omega^{2}+e^{2\psi\left(
t,\rho\right)  }dz^{2},\label{0.0}%
\end{equation}
where $d\Omega^{2}=d\theta_{1}^{2}+\sin^{2}\theta_{1}d\theta_{2}^{2}+\sin
^{2}\theta_{1}\sin^{2}\theta_{2}d\phi^{2}$.

With the metric (\ref{0.0}), the vacuum Einstein equation $R_{ij}-\frac{1}%
{2}g_{ij}R=0$ reads
\begin{equation}
-2\frac{\partial\Phi}{\partial t}\frac{\partial\psi}{\partial t}%
-2\frac{\partial\Phi}{\partial\rho}\frac{\partial\psi}{\partial\rho}+\frac
{3}{\rho}\frac{\partial\Phi}{\partial\rho}+2\frac{\partial^{2}\psi}%
{\partial\rho^{2}}-4\left(  \frac{\partial\psi}{\partial\rho}\right)
^{2}+\frac{9}{\rho}\frac{\partial\psi}{\partial\rho}+\frac{3}{\rho^{2}}\left[
e^{2(\Phi+\psi)}-1\right]  =0,\label{2.1}%
\end{equation}%
\begin{equation}
-2\frac{\partial\Phi}{\partial t}\frac{\partial\psi}{\partial\rho}+\frac
{3}{\rho}\frac{\partial\Phi}{\partial t}-2\frac{\partial\Phi}{\partial\rho
}\frac{\partial\psi}{\partial t}+2\frac{\partial^{2}\psi}{\partial
t\partial\rho}-4\frac{\partial\psi}{\partial t}\frac{\partial\psi}%
{\partial\rho}+\frac{3}{\rho}\frac{\partial\psi}{\partial t}=0,\label{2.2}%
\end{equation}%
\begin{equation}
-2\frac{\partial\Phi}{\partial t}\frac{\partial\psi}{\partial t}%
-2\frac{\partial\Phi}{\partial\rho}\frac{\partial\psi}{\partial\rho}+\frac
{3}{\rho}\frac{\partial\Phi}{\partial\rho}+2\frac{\partial^{2}\psi}{\partial
t^{2}}-4\left(  \frac{\partial\psi}{\partial t}\right)  ^{2}-\frac{3}{\rho
}\frac{\partial\psi}{\partial\rho}-\frac{3}{\rho^{2}}\left[  e^{2(\Phi+\psi
)}-1\right]  =0,\label{2.3}%
\end{equation}%
\begin{equation}
\frac{\partial^{2}\Phi}{\partial\rho^{2}}-\frac{\partial^{2}\Phi}{\partial
t^{2}}+\frac{\partial^{2}\psi}{\partial t^{2}}-\frac{\partial^{2}\psi
}{\partial\rho^{2}}+2\left(  \frac{\partial\psi}{\partial\rho}\right)
^{2}-2\left(  \frac{\partial\psi}{\partial t}\right)  ^{2}-\frac{4}{\rho}%
\frac{\partial\psi}{\partial\rho}-\frac{1}{\rho^{2}}\left[  e^{2(\Phi+\psi
)}-1\right]  =0,\label{2.4}%
\end{equation}%
\begin{equation}
\frac{\partial^{2}\Phi}{\partial\rho^{2}}-\frac{\partial^{2}\Phi}{\partial
t^{2}}+3\frac{\partial^{2}\psi}{\partial t^{2}}-3\frac{\partial^{2}\psi
}{\partial\rho^{2}}+6\left(  \frac{\partial\psi}{\partial\rho}\right)
^{2}-6\left(  \frac{\partial\psi}{\partial t}\right)  ^{2}-\frac{12}{\rho
}\frac{\partial\psi}{\partial\rho}-\frac{3}{\rho^{2}}\left[  e^{2(\Phi+\psi
)}-1\right]  =0.\label{2.5}%
\end{equation}
By light-cone coordinates%
\begin{align}
u &  =t+\rho,\nonumber\\
v &  =t-\rho,
\end{align}
Eqs. (\ref{2.1})-(\ref{2.5}) can be simplified as
\begin{align}
2\left(  -2\frac{\partial\Phi}{\partial u}\frac{\partial\psi}{\partial
u}+\frac{\partial^{2}\psi}{\partial u^{2}}\right)  +3\frac{1}{\rho}\left(
\frac{\partial\Phi}{\partial u}+\frac{\partial\psi}{\partial u}\right)
-4\left(  \frac{\partial\psi}{\partial u}\right)  ^{2} &  =0,\label{2.6}\\
2\left(  -2\frac{\partial\Phi}{\partial v}\frac{\partial\psi}{\partial
v}+\frac{\partial^{2}\psi}{\partial v^{2}}\right)  -3\frac{1}{\rho}\left(
\frac{\partial\Phi}{\partial v}+\frac{\partial\psi}{\partial v}\right)
-4\left(  \frac{\partial\psi}{\partial v}\right)  ^{2} &  =0,\label{2.7}\\
-4\frac{\partial^{2}\psi}{\partial u\partial v}+8\frac{\partial\psi}{\partial
u}\frac{\partial\psi}{\partial v}+3\frac{1}{\rho}\frac{\partial\psi}%
{\partial\rho} &  =0,\label{2.8}\\
\frac{\partial^{2}\Phi}{\partial u\partial v} &  =0,\label{2.9}\\
\frac{\partial\psi}{\partial\rho}+\frac{1}{\rho}\left[  e^{2(\Phi+\psi
)}-1\right]   &  =0.\label{2.10}%
\end{align}
Solving Eq. (\ref{2.10}), we have%
\begin{equation}
\psi=\ln\rho-\frac{1}{2}\ln\left(  2\int_{1}^{\rho}xe^{2\Phi}dx+c_{1}%
(t)\right)  ,\label{2.11}%
\end{equation}
where $c_{1}(t)$ is an arbitrary function of $t$. Substituting Eq.
(\ref{2.11}) into Eq. (\ref{2.8}) gives
\begin{equation}
\frac{d^{2}c_{1}(t)}{dt^{2}}+2\int_{1}^{\rho}2xe^{2\Phi}\left[  2\left(
\frac{\partial\Phi}{\partial t}\right)  ^{2}+\frac{\partial^{2}\Phi}{\partial
t^{2}}\right]  \,dx-4\rho e^{2\Phi}\frac{\partial\Phi}{\partial\rho}=0.
\end{equation}
Taking the derivative with respect to $\rho$, we arrive at an equation of
$\Phi$,%
\begin{equation}
\rho\left(  \frac{\partial^{2}\Phi}{\partial t^{2}}-\frac{\partial^{2}\Phi
}{\partial\rho^{2}}\right)  +2\rho\left[  \left(  \frac{\partial\Phi}{\partial
t}\right)  ^{2}-\left(  \frac{\partial\Phi}{\partial\rho}\right)  ^{2}\right]
-\frac{\partial\Phi}{\partial\rho}=0.\label{2.12}%
\end{equation}
Rewriting Eq. (\ref{2.12}) in light-cone coordinates, we have
\begin{equation}
2\left(  u-v\right)  \frac{\partial^{2}\Phi}{\partial u\partial v}+4\left(
u-v\right)  \frac{\partial\Phi}{\partial u}\frac{\partial\Phi}{\partial
v}-\frac{\partial\Phi}{\partial u}+\frac{\partial\Phi}{\partial v}%
=0.\label{2.14}%
\end{equation}
By Eq. (\ref{2.9}), $\Phi$ can be expressed as%
\begin{equation}
\Phi=f_{1}\left(  u\right)  +f_{2}\left(  v\right)  .\label{2.15}%
\end{equation}
Substituting Eq. (\ref{2.15}) into Eq. (\ref{2.14}) and separating variables,
we have%
\begin{equation}
\frac{1}{df_{1}\left(  u\right)  /du}+4u=\frac{1}{df_{2}\left(  v\right)
/dv}+4v=4a,\label{2.17}%
\end{equation}
where $a$ is the separation constant. Then we arrive at%
\begin{align}
f_{1}\left(  u\right)   &  =-\frac{1}{4}\ln\left(  u+a\right)  =-\frac{1}%
{4}\ln\left(  t+\rho+a\right)  ,\label{2.18}\\
f_{2}\left(  v\right)   &  =-\frac{1}{4}\ln\left(  v+a\right)  +C_{0}%
=-\frac{1}{4}\ln\left(  t-\rho+a\right)  +C_{0}.\label{2.19}%
\end{align}
By Eq. (\ref{2.15}) and then by Eq. (\ref{2.11}), we obtain%
\begin{align}
\Phi &  =-\frac{1}{4}\ln\left(  t+\rho+a\right)  -\frac{1}{4}\ln\left(
t-\rho+a\right)  +C_{0},\label{2.22}\\
\psi &  =\ln\rho-\frac{1}{2}\ln\left(  -2e^{2C_{0}}\sqrt{t+\rho+a}\sqrt
{t-\rho+a}+c_{1}(t)\right)  .\label{2.20}%
\end{align}
$c_{1}(t)$ in Eq. (\ref{2.20}) can be determined by substituting Eq.
(\ref{2.20}) into Eq. (\ref{2.8}). This gives $d^{2}c_{1}(t)/dt^{2}=0$ and
then
\begin{equation}
c_{1}\left(  t\right)  =\alpha t+\beta,\label{ex.1}%
\end{equation}
where $\alpha$ and $\beta$ are constants. Substituting Eqs. (\ref{2.22}%
)-(\ref{ex.1}) into Eqs. (\ref{2.1}) and (\ref{2.2}) gives%
\begin{equation}
\beta=a\alpha.
\end{equation}
Then we obtain the solution%
\begin{align}
\Phi &  =-\frac{1}{4}\ln\left(  t+\rho+a\right)  -\frac{1}{4}\ln\left(
t-\rho+a\right)  +C_{0},\label{2.23}\\
\psi &  =\ln\rho-\frac{1}{2}\ln\left(  -2e^{2C_{0}}\sqrt{t+\rho+a}\sqrt
{t-\rho+a}+\alpha t+a\alpha\right)  .\label{2.24}%
\end{align}

Redefining the variables as $\alpha\left(  t+a\right)  \rightarrow t$,
$\alpha\rho\rightarrow\rho$, $z/\alpha\rightarrow z$, and $1+M=2e^{2C_{0}%
}/\alpha$, we have%
\begin{align}
\Phi &  =-\frac{1}{4}\ln\left(  t^{2}-\rho^{2}\right)  +\frac{1}{2}\ln
\frac{1+M}{2},\label{3.1}\\
\psi &  =\ln\rho-\frac{1}{2}\ln\left[  t-\left(  1+M\right)  \sqrt{t^{2}%
-\rho^{2}}\right]  . \label{3.2}%
\end{align}

The metric then reads%
\begin{equation}
ds^{2}=\frac{1+M}{2\sqrt{t^{2}-\rho^{2}}}\left(  -dt^{2}+d\rho^{2}\right)
+\left[  t-\left(  1+M\right)  \sqrt{t^{2}-\rho^{2}}\right]  d\Omega^{2}%
+\frac{\rho^{2}dz^{2}}{t-\left(  1+M\right)  \sqrt{t^{2}-\rho^{2}}},
\label{3.3}%
\end{equation}
where $t\geq\rho\geq0$ and the parameter $M$ is an arbitrary constant.

When $M=0$, the spacetime described by the metric (\ref{3.3}) reduces to a
flat spacetime. Concretely, we perform a coordinate transformation
\begin{align}
t &  =\frac{1}{2}\left(  \chi^{2}-\xi^{2}+\varrho^{2}\right)  ,\label{3.5}\\
\rho &  =\varrho\sqrt{\chi^{2}-\xi^{2}},\label{3.6}\\
z &  =\frac{1}{2}\ln\frac{\chi+\xi}{\chi-\xi},\label{3.7}%
\end{align}
(the inverse transformation is $\chi=\sqrt{t+\sqrt{t^{2}-\rho^{2}}}\cosh z$,
$\xi=\sqrt{t+\sqrt{t^{2}-\rho^{2}}}\sinh z$, and $\varrho=\sqrt{t-\sqrt
{t^{2}-\rho^{2}}}$), the metric (\ref{3.3}) becomes
\begin{align}
ds^{2} &  =-\frac{1+M}{\chi^{2}-\xi^{2}}\left(  \chi d\chi-\xi d\xi\right)
^{2}+\left(  1+M\right)  d\varrho^{2}+\left[  \varrho^{2}-\frac{M}{2}\left(
\chi^{2}-\xi^{2}-\varrho^{2}\right)  \right]  d\Omega^{2}\nonumber\\
&  +\frac{\varrho^{2}}{\left[  \varrho^{2}-\frac{M}{2}\left(  \chi^{2}-\xi
^{2}-\varrho^{2}\right)  \right]  \left(  \chi^{2}-\xi^{2}\right)  }\left(
\chi d\xi-\xi d\chi\right)  ^{2}.
\end{align}
Taking $M=0$, it reduces to $ds^{2}=-d\chi^{2}+d\xi^{2}+d\varrho^{2}%
+\varrho^{2}d\Omega^{2}$.

\section{Singularities of the spacetime \label{singularity}}

In this section, we analyze the singularity of the spacetime described by the
metric (\ref{3.3}).

The metric (\ref{3.3}) is singular at%
\begin{align}
\rho &  =t,\label{3.11}\\
\rho &  =\frac{\sqrt{M\left(  M+2\right)  }}{M+1}t. \label{3.12}%
\end{align}
In order to show the curvature singularity and the scalar polynomial curvature
singularity, we here calculate the Riemann tensor $R_{\text{ }ikl}^{j}$ and
the scalar $R_{\text{ }ikl}^{j}R_{j}^{\text{ }ikl}$ in the orthonormal frame.

An orthonormal frame $\left(  e_{0},e_{1},e_{2},\ldots,e_{m}\right)  $ and a
natural frame $\left(  \frac{\partial}{\partial x^{0}},\frac{\partial
}{\partial x^{1}},\frac{\partial}{\partial x^{2}},\ldots,\frac{\partial
}{\partial x^{m}}\right)  $ on a manifold are connected by
\begin{equation}
e_{i}=X_{i}^{k}\frac{\partial}{\partial x^{k}},
\end{equation}
where $X_{i}^{k}$ is a nondegenerate matrix. Choosing the corresponding
$1$-form
\begin{equation}
\theta^{i}=X_{j}^{\ast i}dx^{j}%
\end{equation}
with $X_{j}^{\ast i}$ the inverse matrix of $X_{i}^{k}$, we can express the
metric of the manifold as
\begin{equation}
ds^{2}=g_{ij}dx^{i}dx^{j}=-\left(  \theta^{0}\right)  ^{2}+\sum_{i=1}%
^{n}\left(  \theta^{i}\right)  ^{2}.
\end{equation}

In the orthonormal frame, the curvature components $R_{\text{ }ikl}^{j}$ can
be obtained from the curvature components in the natural frame, $\mathcal{R}%
_{\text{ }ikl}^{j}$, by%

\begin{align}
R  &  =R_{\text{ }ikl}^{j}\theta^{i}\otimes e_{j}\otimes\theta^{k}%
\otimes\theta^{l}\nonumber\\
&  =\mathcal{R}_{\text{ }ikl}^{j}dx^{i}\otimes\frac{\partial}{\partial x^{j}%
}\otimes dx^{k}\otimes dx^{l}.
\end{align}
Then we arrive at \cite{lam2000lectures}%
\begin{equation}
R_{\text{ }ikl}^{j}=X_{q}^{\ast j}X_{i}^{p}X_{k}^{r}X_{l}^{s}\mathcal{R}%
_{\text{ }prs}^{q}.
\end{equation}
The curvature components $R_{\text{ }ikl}^{j}$ are independent of the coordinates.

The components of the Riemann tensor $R_{\text{ }ikl}^{j}$ in the orthonormal
frame can be calculated from the metric (\ref{3.3}) directly:%
\begin{align}
R_{\text{ }220}^{0} &  =R_{\text{ }330}^{0}=R_{\text{ }440}^{0}=\rho
^{2}\emph{U}\left(  t,\rho\right)  \emph{V}\left(  t,\rho\right)  ,\nonumber\\
R_{\text{ }221}^{1} &  =R_{\text{ }331}^{1}=R_{\text{ }441}^{1}=-t^{2}%
\emph{U}\left(  t,\rho\right)  \emph{V}\left(  t,\rho\right)  ,\nonumber\\
R_{\text{ }220}^{1} &  =R_{\text{ }330}^{1}=R_{\text{ }440}^{1}=t\rho
\emph{U}\left(  t,\rho\right)  \emph{V}\left(  t,\rho\right)  ,\nonumber\\
R_{\text{ }550}^{0} &  =-3\rho^{2}\emph{U}\left(  t,\rho\right)
\emph{V}\left(  t,\rho\right)  ,\text{ }\nonumber\\
R_{\text{ }551}^{1} &  =3t^{2}\emph{U}\left(  t,\rho\right)  \emph{V}\left(
t,\rho\right)  ,\nonumber\\
R_{\text{ }550}^{1} &  =-3t\rho\emph{U}\left(  t,\rho\right)  \emph{V}\left(
t,\rho\right)  ,\nonumber\\
R_{\text{ }332}^{2} &  =R_{\text{ }442}^{2}=-R_{\text{ }552}^{2}=-R_{\text{
}443}^{3}=-R_{\text{ }553}^{3}=-R_{\text{ }554}^{4}=\sqrt{t^{2}-\rho^{2}%
}\emph{U}\left(  t,\rho\right)  ;
\end{align}
the scalar then becomes
\begin{equation}
R_{\text{ }ikl}^{j}R_{j}^{\text{ }ikl}=72\left(  t^{2}-\rho^{2}\right)
\emph{U}^{2}\left(  t,\rho\right)  .
\end{equation}
Here, the singularities (\ref{3.11}) and (\ref{3.12}) are embodied in the
factors
\begin{align}
\emph{U}\left(  t,\rho\right)   &  =\frac{M(M+2)}{2(M+1)}\frac{1}{\left[
t-\left(  1+M\right)  \sqrt{t^{2}-\rho^{2}}\right]  ^{2}},\\
\emph{V}\left(  t,\rho\right)   &  =\frac{1}{\sqrt{t^{2}-\rho^{2}}},
\end{align}
respectively.

Recall that the singularity of the components of the Riemann tensor $R_{\text{
}ikl}^{j}$ is the curvature singularity and the singularity of the scalar
$R_{\text{ }ikl}^{j}R_{j}^{\text{ }ikl}$ is the scalar polynomial curvature
singularity \cite{hawking1974large,wald2010general}. Now we see that (1) the
singularity $\rho=t$ in the metric are the curvature singularities; (2) the
singularity $\rho=\frac{\sqrt{M\left(  M+2\right)  }}{M+1}t$ are both the
curvature singularity and the scalar polynomial curvature singularity.

\section{Extension of the spacetime: removing singularity \label{extension}}

In this section, we make an extension of the spacetime described by the metric
(\ref{3.3}). It will be seen that the curvature singularity in the extended
spacetime corresponding to $\rho=t$ in the metric (\ref{3.3}) is no longer
singular; while the scalar polynomial curvature singularity in the extended
spacetime corresponding to $\rho=\frac{\sqrt{M\left(  M+2\right)  }}{M+1}t$ is
still singular. That is to say, some curvature singularities can be removed.

Introduce new coordinates $\left(  \eta,r\right)  $ by%
\begin{align}
t  &  =\frac{1}{2}\left(  \eta^{2}+r^{2}\right)  ,\nonumber\\
\rho &  =\eta r, \label{C.3}%
\end{align}
where $\eta\geq r\geq0$. In the coordinates $\left(  \eta,r\right)  $, the
metric (\ref{3.3}) becomes%
\begin{equation}
ds^{2}=\left(  1+M\right)  \left(  -d\eta^{2}+dr^{2}\right)  +\frac{1}%
{2}\left[  \left(  2+M\right)  r^{2}-M\eta^{2}\right]  d\Omega^{2}+\frac
{2\eta^{2}r^{2}dz^{2}}{\left(  2+M\right)  r^{2}-M\eta^{2}}. \label{C.4}%
\end{equation}
The metric (\ref{C.4}) with $\eta\geq0$ and $r\geq0$ describes a larger
spacetime than the spacetime described by the metric (\ref{3.3}). In fact, the
spacetime described by the metric (\ref{3.3}) is isometric to the area
$\eta\geq r\geq0$ in the metric (\ref{C.4}). That is, the spacetime described
by the metric (\ref{C.4}) is an extension of the solution (\ref{3.3}).

The metric (\ref{C.4}) is singular at
\begin{equation}
r=\sqrt{\frac{M}{M+2}}\eta. \label{C.5}%
\end{equation}
Similarly, we can calculate the Riemann tensor $R_{\text{ }ikl}^{j}$ and the
scalar $R_{\text{ }ikl}^{j}R_{j}^{\text{ }ikl}$ from the metric (\ref{C.4}) in
the orthonormal frame.

The Riemann tensor components in the orthonormal frame are
\begin{align}
R_{\text{ }220}^{0} &  =R_{\text{ }330}^{0}=R_{\text{ }440}^{0}=r^{2}%
\emph{W}\left(  \eta,r\right)  ,\text{ }\nonumber\\
R_{\text{ }220}^{1} &  =R_{\text{ }330}^{1}=R_{\text{ }440}^{1}=\eta
r\emph{W}\left(  \eta,r\right)  ,\nonumber\\
R_{\text{ }221}^{1} &  =R_{\text{ }331}^{1}=R_{\text{ }441}^{1}=-\eta
^{2}\emph{W}\left(  \eta,r\right)  ,\nonumber\\
R_{\text{ }550}^{0} &  =-3r^{2}\emph{W}\left(  \eta,r\right)  ,\text{
}R_{\text{ }551}^{1}=3\eta^{2}\emph{W}\left(  \eta,r\right)  ,\text{
}R_{\text{ }551}^{0}=3\eta r\emph{W}\left(  \eta,r\right)  ,\nonumber\\
R_{\text{ }332}^{2} &  =R_{\text{ }442}^{2}=-R_{\text{ }552}^{2}=R_{\text{
}443}^{3}=-R_{\text{ }553}^{3}=-R_{\text{ }554}^{4}=\left(  \eta^{2}%
-r^{2}\right)  \emph{W}\left(  \eta,r\right)  ;
\end{align}
the scalar is%
\begin{equation}
R_{\text{ }ikl}^{j}R_{j}^{\text{ }ikl}=72\left(  \eta^{2}-r^{2}\right)
^{2}\emph{W}^{2}\left(  \eta,r\right)  .\label{C.9}%
\end{equation}
After the extension, there is only one singularity (\ref{C.5}) remaining in
the curvature, embodied in the factor%

\begin{equation}
\emph{W}\left(  \eta,r\right)  =\frac{M(M+2)}{2(M+1)}\frac{1}{\left[
(M+2)r^{2}-M\eta^{2}\right]  ^{2}}.
\end{equation}

Analyzing the components of the Riemann tensor $R_{\text{ }ikl}^{j}$ and the
scalar $R_{\text{ }ikl}^{j}R_{j}^{\text{ }ikl}$, we see that $\eta=r$
corresponding to $\rho=t$ in the metric (\ref{3.3}) is no longer singular, or,
the curvature singularity at $\rho=t$ is removed by the extension. Concretely,
we extend the $1+5$-dimensional solution (\ref{3.3}) to a larger spacetime
through the coordinates transform (\ref{C.3}). This extended spacetime is
described by the metric (\ref{C.4}). The scalar $R_{\text{ }ikl}^{j}%
R_{j}^{\text{ }ikl}$ with the metric (\ref{C.4}) at $r=\sqrt{\frac{M}{M+2}%
}\eta$ corresponding to $\rho=\frac{\sqrt{M\left(  M+2\right)  }}{M+1}t$ in
the metric (\ref{3.3}) is still singular.

\section{Singularity hypersurface \label{hypersurface}}

The singularity in the spacetime described by the metric (\ref{3.3}) is not
zero dimensional points but two hypersurfaces.

The singularity $\rho=\frac{\sqrt{M\left(  M+2\right)  }}{M+1}t$ is a
spacetime singularity, since the components of Riemann tensor $R_{\text{ }%
ikl}^{j}$ and the scalar $R_{\text{ }ikl}^{j}R_{j}^{\text{ }ikl}$ blows up.
When $\rho=\frac{\sqrt{M\left(  M+2\right)  }}{M+1}t$, the metric (\ref{3.3})
describes a hypersurface.

It is worth mentioning that the points determined by $\rho=\frac
{\sqrt{M\left(  M+2\right)  }}{M+1}t$ form a hypersurface behaving like a
horizon. This can be seen from the following two facts.

1. The velocity of a light signal propagating in the $z$ direction is
\begin{equation}
\frac{dz}{dt}=\pm\sqrt{\frac{1+M}{2}}\frac{\sqrt{\left\vert t-\left(
1+M\right)  \sqrt{t^{2}-\rho^{2}}\right\vert }}{\rho\left(  t^{2}-\rho
^{2}\right)  ^{1/4}}.
\end{equation}
When $\rho=\frac{\sqrt{M\left(  M+2\right)  }}{M+1}t$, the coordinate velocity
of the light signal vanishes. Similarly, the coordinate velocity of a light
signal in the Schwarzschild spacetime in the radial direction is
$dr/dt=\pm\left(  1-\frac{2M}{r}\right)  $ which also vanishes at $r=2M$
\cite{ohanian2013gravitation}.

2. In the region $0<\rho<\frac{\sqrt{M\left(  M+2\right)  }}{M+1}t$, the signs
of $g_{ii}$ ($i=0,1,2,3,4,5$) become the opposite. Concretely, when
$\frac{\sqrt{M\left(  M+2\right)  }}{M+1}t<\rho<t$, the coordinates $z$,
$\theta_{1}$, $\theta_{2}$, and $\phi$ are spacelike, while when $0<\rho
<\frac{\sqrt{M\left(  M+2\right)  }}{M+1}t$, the coordinates $z$, $\theta_{1}%
$, $\theta_{2}$, and $\phi$ are timelike. In other words, in the region
$\frac{\sqrt{M\left(  M+2\right)  }}{M+1}t<\rho<t$, the metric normal form is
$\operatorname{diag}\left(  -,+,+,+,+,+\right)  $, while in the region
$0<\rho<\frac{\sqrt{M\left(  M+2\right)  }}{M+1}t$, the metric normal form is
$\operatorname{diag}\left(  -,+,-,-,-,-\right)  $.

Nevertheless, this singularity hypersurface is not a horizon like that in the
Schwarzschild spacetime, since it is not a one-way membrane.

\section{Gravitational wave: energy-momentum pseudotensor
\label{gravitational wave}}

In this section, we show that the $1+5$-dimensional solution (\ref{3.3}) is a
gravitational-wave solution.

In the metric (\ref{3.3}), the hypersurface $\rho=t$ can be regarded as the
boundary of the spacetime. It provides a picture that the scale of the
spacetime expands with time going on. To illustrate that this solution
describes a gravitational wave, we will show that the solution has a
time-varying energy-momentum pseudotensor (energy-momentum complex). The
energy-momentum pseudotensor is defined to describe the energy-momentum of the
gravitational field. The reason why using pseudotensors is that according to
the equivalence principle, the spacetime is flat at a given point in local
geodesic coordinates; as a result the energy-momentum of a gravitational field
in local geodesic coordinates vanishes.\textbf{ }If a gravitational field has
a time-varying energy-momentum pseudotensor, the solution may represent a
gravitational wave
\cite{virbhadra1995energy,rosen1993energy,landau2013classical}. More
discussions on energy-momentum pseudotensors can be found in Ref.
\cite{chang1999pseudotensors}.

The definition of the energy-momentum pseudotensor is not unanimous. There are
many different definitions for energy-momentum pseudotensors:\ the Einstein
energy-momentum pseudotensor $\theta_{i}^{j}=\frac{1}{16\pi}H_{i\text{ \ }%
,k}^{jk}$ with $H_{i}^{jk}=\frac{1}{\sqrt{-g}}g_{il}[-g(g^{jl}g^{km}%
-g^{kl}g^{jm})]_{,m}$ \cite{rosen1993energy}; the Landau-Lifshitz
energy-momentum pseudotensor $L^{mn}=\frac{1}{16\pi}S_{\text{ \ \ \ }%
,jk}^{mjnk}$ with $S^{mjnk}=-g(g^{mn}g^{jk}-g^{mk}g^{jn})$
\cite{landau2013classical}; the Tolman energy-momentum pseudotensor
$\mathcal{T}_{k}^{i}=\frac{1}{8\pi}U_{k,j}^{ij}$ with $U_{k}^{ij}=\sqrt
{-g}[-g^{li}(-\Gamma_{kl}^{i}+\frac{1}{2}g_{k}^{i}\Gamma_{al}^{a}+\frac{1}%
{2}g_{l}^{j}\Gamma_{al}^{a})+\frac{1}{2}g_{k}^{i}g^{lm}(-\Gamma_{lm}^{j}%
+\frac{1}{2}g_{l}^{j}\Gamma_{am}^{a}+\frac{1}{2}g_{m}^{j}\Gamma_{al}^{a})]$
\cite{virbhadra1995energy}; the Papapetrou energy-momentum pseudotensor
$\Omega^{ij}=\frac{1}{16\pi}\mathcal{N}_{\text{ \ \ },ab}^{ikab}$ with
$\mathcal{N}^{ikab}=\sqrt{-g}(g^{ik}\eta^{ab}-g^{ia}\eta^{kb}+g^{ab}\eta
^{ik}-g^{kb}\eta^{ia})$ \cite{papapetrou1948static}; the Weinberg
energy-momentum pseudotensor $t_{ij}=\frac{1}{8\pi}(R_{ij}-\frac{1}{2}%
g_{ij}R-R_{ij}^{\left(  1\right)  }+\frac{1}{2}\eta_{ij}R_{\text{ }%
k}^{k\left(  1\right)  })$ with $R_{ij}^{\left(  1\right)  }=\frac{1}{2}%
\eta^{kl}(\frac{\partial^{2}h_{lk}}{\partial x^{i}\partial x^{j}}%
-\frac{\partial^{2}h_{li}}{\partial x^{ik}\partial x^{j}}-\frac{\partial
^{2}h_{lj}}{\partial x^{k}\partial x^{i}}+\frac{\partial^{2}h_{ij}}{\partial
x^{k}\partial x^{l}})$ and\ $g_{ij}=\eta_{ij}+h_{ij}$
\cite{weinberg1972gravitation}; the M\O ller energy-momentum pseudotensor
$\mathcal{J}_{j}^{i}=\frac{1}{8\pi}\xi_{j}^{ik}{}_{,k}$ with $\xi_{j}%
^{ik}=\sqrt{-g}(\frac{\partial g_{jl}}{\partial x^{m}}-\frac{\partial g_{im}%
}{\partial x^{l}})g^{im}g^{kl}$ \cite{moller1958localization}.

Direct calculation shows that, for the $1+5$-dimensional solution (\ref{3.3}),
all the energy-momentum pseudotensors mentioned above in the "Cartesian
coordinates" \cite{virbhadra1995energy}
\begin{align}
x_{1} &  =\rho\sin\theta_{1}\sin\theta_{2}\sin\phi,\nonumber\\
x_{2} &  =\rho\sin\theta_{1}\sin\theta_{2}\cos\phi,\nonumber\\
x_{3} &  =\rho\sin\theta_{1}\cos\theta_{2},\nonumber\\
x_{4} &  =\rho\cos\theta_{1},\label{Cart.1}%
\end{align}
have time-varying $tt$-components and $tx_{1}$-components.

Additionally, for the extended spacetime (\ref{C.4}), these energy-momentum
pseudotensors in "Cartesian coordinates"
\begin{align}
y_{1}  &  =r\sin\theta_{1}\sin\theta_{2}\sin\phi,\nonumber\\
y_{2}  &  =r\sin\theta_{1}\sin\theta_{2}\cos\phi,\nonumber\\
y_{3}  &  =r\sin\theta_{1}\cos\theta_{2},\nonumber\\
y_{4}  &  =r\cos\theta_{1},
\end{align}
except the M\O ller energy-momentum pseudotensor, also have time-varying
$\eta\eta$-components and $\eta y_{1}$-components. For the M\O ller
energy-momentum pseudotensor, the $\eta\eta$-component vanishes but the $\eta
y_{1}$-component is also time-varying.

That is, the $1+5$-dimensional solution (\ref{3.3}) can be regarded as a
gravitational-wave solution.

\section{Note on $1+n$-dimensional solution \label{1pn}}

In the above, we consider a $1+5$-dimensional solution. In this section, we
wish here to add a few words to the $1+n$-dimensional solutions.

The $1+n$-dimensional cylindrical metric reads
\begin{equation}
ds^{2}=e^{2\Phi\left(  t,\rho\right)  }\left(  -dt^{2}+d\rho^{2}\right)
+e^{-2\psi\left(  t,\rho\right)  }\rho^{2}\sum_{k=1}^{n-1}d\theta_{k}^{2}%
\prod_{s=1}^{k-1}\sin^{2}\theta_{s}+e^{2\psi\left(  t,\rho\right)  }%
dz^{2}.\label{E.1}%
\end{equation}
When $n=2$, the metric becomes $ds^{2}=e^{2\Phi\left(  t,\rho\right)  }\left(
-dt^{2}+d\rho^{2}\right)  +e^{2\psi\left(  t,\rho\right)  }dz^{2}$ and have
only trivial solutions which can be transformed to a Minkowski solution. When
$n\geq3$, the Einstein equation of the metric (\ref{E.1}) in the light-cone
coordinates $u\equiv t+\rho$ and $v\equiv t-\rho$ reads
\begin{align}
\left(  n-3\right)  \left(  -2\frac{\partial\Phi}{\partial u}\frac
{\partial\psi}{\partial u}+\frac{\partial^{2}\psi}{\partial u^{2}}\right)
+\left(  n-2\right)  \frac{1}{\rho}\left(  \frac{\partial\Phi}{\partial
u}+\frac{\partial\psi}{\partial u}\right)  -\left(  n-1\right)  \left(
\frac{\partial\psi}{\partial u}\right)  ^{2} &  =0,\label{N.1}\\
\left(  n-3\right)  \left(  -2\frac{\partial\Phi}{\partial v}\frac
{\partial\psi}{\partial v}+\frac{\partial^{2}\psi}{\partial v^{2}}\right)
-\left(  n-2\right)  \frac{1}{\rho}\left(  \frac{\partial\Phi}{\partial
v}+\frac{\partial\psi}{\partial v}\right)  -\left(  n-1\right)  \left(
\frac{\partial\psi}{\partial v}\right)  ^{2} &  =0,\label{N.2}\\
-\frac{\partial^{2}\Phi}{\partial u\partial v}+\frac{\left(  n-5\right)  }%
{2}\left(  \frac{\partial^{2}\psi}{\partial u\partial v}+\frac{\partial\psi
}{\partial u}\frac{\partial\psi}{\partial v}\right)   &  =0,\label{N.3}\\
-4\frac{\partial^{2}\psi}{\partial u\partial v}+4\left(  n-3\right)
\frac{\partial\psi}{\partial u}\frac{\partial\psi}{\partial v}+\left(
n-2\right)  \frac{1}{\rho}\frac{\partial\psi}{\partial\rho} &  =0,\label{N.4}%
\\
\frac{1}{\rho}\frac{\partial\psi}{\partial\rho}+\frac{1}{\rho^{2}}\left[
e^{2(\Phi+\psi)}-1\right]   &  =0.\label{N.5}%
\end{align}

The $1+5$-dimensional case is special. When $n=5$, Eq. (\ref{N.3}) becomes%
\begin{equation}
\frac{\partial^{2}\Phi}{\partial u\partial v}=0.\label{E.3}%
\end{equation}
The solution of Eq. (\ref{E.3}) then reads, as expressed in Eq. (\ref{2.15}),
$\Phi=F_{1}\left(  u\right)  +F_{2}\left(  v\right)  $. It can be shown that
once Eq. (\ref{E.3}) is satisfied, the spacetime can be extended by
introducing new light-cone coordinates by
\begin{align}
d\tilde{u} &  =e^{2F_{1}}du,\label{E.4}\\
d\tilde{v} &  =e^{2F_{2}}dv.\nonumber
\end{align}
Then in the new coordinates $\left(  \tilde{u},\tilde{v}\right)  $, the metric
(\ref{E.1}) becomes
\begin{equation}
ds^{2}=-d\tilde{u}d\tilde{v}+e^{-2\psi}\rho^{2}\sum_{k=1}^{n-1}d\theta_{k}%
^{2}\prod_{s=1}^{k-1}\sin^{2}\theta_{s}+e^{2\psi}dz^{2}.\label{E.5}%
\end{equation}
The metric (\ref{E.5}) is usually the extension of the metric (\ref{E.1}).

\section{Conclusions and outlook \label{Conclusion}}

In this paper, we show that the curvature singularity is not always a
spacetime singularity. For exemplifying this, we first solve a vacuum solution
of the Einstein equation, which is a $1+5$-dimensional cylindrical
gravitational-wave solution.

We first show that there are two singularities in the Riemann tensor
$R_{\text{ }ikl}^{j}$ and the scalar $R_{\text{ }ikl}^{j}R_{j}^{\text{ }ikl}$
of the $1+5$-dimensional gravitational-wave solution obtained in the present
paper. In other words, there are two curvature singularities in this
spacetime. Then we show that one of these two singularities can be removed by
an extension of the spacetime. That is to say, the curvature singularity which
can be removed by the extension is not a real spacetime singularity, or, the
curvature singularity cannot be used as a criterion for spacetime
singularities. We also show that there is a horizon-like singularity hypersurface.

Moreover, in the above, we only consider the case of $M>0$. When $M<0$, the
situation is different. In this case, when $M=-2$, the metric (\ref{C.4})
reduces to a Minkowski spacetime described by $ds^{2}=d\eta^{2}-dr^{2}%
+\eta^{2}d\Omega^{2}+r^{2}dz^{2}$.

The spacetime described by the metric (\ref{3.3}) has only one curvature
singularity at $\rho=t$. When $-1<M<0$, the metric normal form is
$\operatorname{diag}\left(  -,+,+,+,+,+\right)  $ and when $M<-1$ ($M\neq-2$)
the metric normal form is $\operatorname{diag}\left(  +,-,+,+,+,+\right)  $.

After the extension of the spacetime by the transformation of coordinates
(\ref{C.3}), when $-2<M<0$ ($M\neq-1$), there is no singularity. The metric
normal form when $-1<M<0$ is $\operatorname{diag}\left(  -,+,+,+,+,+\right)  $
and the metric normal form when $-2<M<-1$ is $\operatorname{diag}\left(
+,-,+,+,+,+\right)  $. When $M<-2$, there is a scalar polynomial curvature
singularity at $r=\sqrt{\frac{M}{M+2}}\eta$. In the region $r>\sqrt{\frac
{M}{M+2}}\eta$, the metric normal form is $\operatorname{diag}\left(
+,-,+,+,+,+\right)  $ and in the region $r<\sqrt{\frac{M}{M+2}}\eta$, the
metric normal form is $\operatorname{diag}\left(  +,-,-,-,-,-\right)  $.

In future studies, we can start from the solution given in the present paper
to consider the influence of the singularity to quantum effects by
calculating, e.g., the partition function and the one-loop effective action
based on the heat-kernel method \cite{dai2009number,dai2010approach}.

The singularity problem is an important problem in gravity theory. Some
authors consider the singularity theorem under more general conditions. The
Raychaudhuri equation and the singularity theorem in the Finsler spacetime are
studied in Ref. \cite{aazami2015penrose,minguzzi2015raychaudhuri}. Some
singularity theorems are proved in $\mathbf{C}^{1,1}$ metrics
\cite{kunzinger2015hawking,kunzinger2015penrose}. There are also some analyses
of the singularity starting from exact solutions of the Einstein equation
\cite{sarma2016axially,ahmed2017cylindrically,olmo2015geodesic,mkenyeleye2014gravitational,stoica2012schwarzschild}%
. It is shown that the central singularity can be replaced by a bounce by
taking care of the quantum effects in the inhomogeneous dust collapse
\cite{liu2014singularity}.


\acknowledgments

We are very indebted to Dr G. Zeitrauman for his encouragement. This work is supported in part by NSF of China under Grant
No. 11575125 and No. 11675119.










\providecommand{\href}[2]{#2}\begingroup\raggedright\endgroup


\begin{thebibliography}{10}

\bibitem{wald2010general}
R.~M. Wald, {\em General relativity}.
\newblock University of Chicago press, 2010.

\bibitem{geroch1968singularity}
R.~Geroch, {\it What is a singularity in general relativity?},  {\em Annals of
  Physics} {\bf 48} (1968), no.~3 526--540.

\bibitem{geroch1968local}
R.~Geroch, {\it Local characterization of singularities in general relativity},
   {\em Journal of Mathematical Physics} {\bf 9} (1968), no.~3 450--465.

\bibitem{schmidt1971new}
B.~Schmidt, {\it A new definition of singular points in general relativity},
  {\em General relativity and gravitation} {\bf 1} (1971), no.~3 269--280.

\bibitem{johnson1977bundle}
R.~A. Johnson, {\it The bundle boundary in some special cases},  {\em Journal
  of Mathematical Physics} {\bf 18} (1977), no.~5 898--902.

\bibitem{geroch1982singular}
R.~Geroch, L.~Can-bin, and R.~M. Wald, {\it Singular boundaries of
  space--times},  {\em Journal of Mathematical Physics} {\bf 23} (1982), no.~3
  432--435.

\bibitem{hawking1974large}
S.~W. Hawking, G.~F.~R. Ellis, and R.~Sachs, {\it The large scale structure of
  space-time},  1974.

\bibitem{lam2000lectures}
S.~C.-H. C.-K. Lam, S.~Chern, and W.~Chen, {\it Lectures on differential
  geometry},  2000.

\bibitem{misner1973gravitation}
C.~W. Misner, K.~S. Thorne, and J.~A. Wheeler, {\em Gravitation}.
\newblock Macmillan, 1973.

\bibitem{landau2013classical}
L.~D. Landau, {\em The classical theory of fields}, vol.~2.
\newblock Elsevier, 2013.

\bibitem{sathyaprakash2009physics}
B.~S. Sathyaprakash and B.~F. Schutz, {\it Physics, astrophysics and cosmology
  with gravitational waves},  {\em Living Reviews in Relativity} {\bf 12}
  (2009), no.~1 2.

\bibitem{abbott2016observation}
B.~P. Abbott, R.~Abbott, T.~Abbott, M.~Abernathy, F.~Acernese, K.~Ackley,
  C.~Adams, T.~Adams, P.~Addesso, R.~Adhikari, et~al., {\it Observation of
  gravitational waves from a binary black hole merger},  {\em Physical review
  letters} {\bf 116} (2016), no.~6 061102.

\bibitem{aasi2014gravitational}
J.~Aasi, J.~Abadie, B.~Abbott, R.~Abbott, T.~Abbott, M.~Abernathy, T.~Accadia,
  F.~Acernese, C.~Adams, T.~Adams, et~al., {\it Gravitational waves from known
  pulsars: results from the initial detector era},  {\em The Astrophysical
  Journal} {\bf 785} (2014), no.~2 119.

\bibitem{centrella2010black}
J.~Centrella, J.~G. Baker, B.~J. Kelly, and J.~R. van Meter, {\it Black-hole
  binaries, gravitational waves, and numerical relativity},  {\em Reviews of
  Modern Physics} {\bf 82} (2010), no.~4 3069.

\bibitem{senatore2014new}
L.~Senatore, E.~Silverstein, and M.~Zaldarriaga, {\it New sources of
  gravitational waves during inflation},  {\em Journal of Cosmology and
  Astroparticle Physics} {\bf 2014} (2014), no.~08 016.

\bibitem{kakizaki2015gravitational}
M.~Kakizaki, S.~Kanemura, and T.~Matsui, {\it Gravitational waves as a probe of
  extended scalar sectors with the first order electroweak phase transition},
  {\em Physical Review D} {\bf 92} (2015), no.~11 115007.

\bibitem{momeni2016cylindrical}
D.~Momeni, K.~Myrzakulov, R.~Myrzakulov, and M.~Raza, {\it Cylindrical
  solutions in mimetic gravity},  {\em The European Physical Journal C} {\bf
  76} (2016), no.~6 301.

\bibitem{delice2013higher}
{\"O}.~Delice, P.~Kirezli, and D.~K. {\c{C}}iftci, {\it Higher dimensional
  cylindrical or kasner type electrovacuum solutions},  {\em General Relativity
  and Gravitation} {\bf 45} (2013), no.~11 2251--2272.

\bibitem{emparan2002rotating}
R.~Emparan and H.~S. Reall, {\it A rotating black ring solution in five
  dimensions},  {\em Physical Review Letters} {\bf 88} (2002), no.~10 101101.

\bibitem{chen2006general}
W.~Chen, H.~L{\"u}, and C.~Pope, {\it General kerr--nut--ads metrics in all
  dimensions},  {\em Classical and Quantum Gravity} {\bf 23} (2006), no.~17
  5323.

\bibitem{rahaman2006thin}
F.~Rahaman, M.~Kalam, and S.~Chakraborty, {\it Thin shell wormholes in higher
  dimensional einstein--maxwell theory},  {\em General Relativity and
  Gravitation} {\bf 38} (2006), no.~11 1687--1695.

\bibitem{aliev2006rotating}
A.~Aliev, {\it Rotating black holes in higher dimensional einstein-maxwell
  gravity},  {\em Physical Review D} {\bf 74} (2006), no.~2 024011.

\bibitem{hollands2007higher}
S.~Hollands, A.~Ishibashi, and R.~M. Wald, {\it A higher dimensional stationary
  rotating black hole must be axisymmetric},  {\em Communications in
  mathematical physics} {\bf 271} (2007), no.~3 699--722.

\bibitem{vacaru2010general}
S.~I. Vacaru, {\it On general solutions for field equations in einstein and
  higher dimension gravity},  {\em International Journal of Theoretical
  Physics} {\bf 49} (2010), no.~4 884--913.

\bibitem{tomizawa2006vacuum1}
S.~Tomizawa, Y.~Morisawa, and Y.~Yasui, {\it Vacuum solutions of five
  dimensional einstein equations generated by inverse scattering method},  {\em
  Physical Review D} {\bf 73} (2006), no.~6 064009.

\bibitem{tomizawa2006vacuum2}
S.~Tomizawa and M.~Nozawa, {\it Vacuum solutions of five dimensional einstein
  equations generated byinverse scattering method. ii. production of the black
  ring solution},  {\em Physical Review D} {\bf 73} (2006), no.~12 124034.

\bibitem{padmanabhan2010gravitation}
T.~Padmanabhan, {\em Gravitation: foundations and frontiers}.
\newblock Cambridge University Press, 2010.

\bibitem{mukohyama2000brane}
S.~Mukohyama, {\it Brane-world solutions, standard cosmology, and dark
  radiation},  {\em Physics Letters B} {\bf 473} (2000), no.~3 241--245.

\bibitem{gherghetta2000living}
T.~Gherghetta, E.~Roessl, and M.~Shaposhnikov, {\it Living inside a hedgehog:
  higher-dimensional solutions that localize gravity},  {\em Physics Letters B}
  {\bf 491} (2000), no.~3 353--361.

\bibitem{padmanabhan2013lanczos}
T.~Padmanabhan and D.~Kothawala, {\it Lanczos--lovelock models of gravity},
  {\em Physics Reports} {\bf 531} (2013), no.~3 115--171.

\bibitem{dotti2007static}
G.~Dotti, J.~Oliva, and R.~Troncoso, {\it Static wormhole solution for
  higher-dimensional gravity in vacuum},  {\em Physical Review D} {\bf 75}
  (2007), no.~2 024002.

\bibitem{ohanian2013gravitation}
H.~C. Ohanian and R.~Ruffini, {\em Gravitation and spacetime}.
\newblock Cambridge University Press, 2013.

\bibitem{virbhadra1995energy}
K.~Virbhadra, {\it Energy and momentum of cylindrical gravitational waves-ii},
  {\em Pramana} {\bf 45} (1995), no.~2 215--219.

\bibitem{rosen1993energy}
N.~Rosen and K.~Virbhadra, {\it Energy and momentum of cylindrical
  gravitational waves},  {\em General Relativity and Gravitation} {\bf 25}
  (1993), no.~4 429--433.

\bibitem{chang1999pseudotensors}
C.-C. Chang, J.~M. Nester, and C.-M. Chen, {\it Pseudotensors and quasilocal
  energy-momentum},  {\em Physical Review Letters} {\bf 83} (1999), no.~10
  1897.

\bibitem{papapetrou1948static}
A.~Papapetrou, {\it Static spherically symmetric solutions in the unitary field
  theory},  in {\em Proceedings of the Royal Irish Academy. Section A:
  Mathematical and Physical Sciences}, vol.~52, pp.~69--86, JSTOR, 1948.

\bibitem{weinberg1972gravitation}
S.~Weinberg, {\em Gravitation and cosmology: principles and applications of the
  general theory of relativity}, vol.~1.
\newblock Wiley New York, 1972.

\bibitem{moller1958localization}
C.~M{\o}ller, {\it On the localization of the energy of a physical system in
  the general theory of relativity},  {\em Annals of Physics} {\bf 4} (1958),
  no.~4 347--371.

\bibitem{dai2009number}
W.-S. Dai and M.~Xie, {\it The number of eigenstates: counting function and
  heat kernel},  {\em Journal of High Energy Physics} {\bf 2009} (2009), no.~02
  033.

\bibitem{dai2010approach}
W.-S. Dai and M.~Xie, {\it An approach for the calculation of one-loop
  effective actions, vacuum energies, and spectral counting functions},  {\em
  Journal of High Energy Physics} {\bf 2010} (2010), no.~6 1--29.

\bibitem{aazami2015penrose}
A.~B. Aazami and M.~A. Javaloyes, {\it Penrose's singularity theorem in a
  finsler spacetime},  {\em Classical and Quantum Gravity} {\bf 33} (2015),
  no.~2 025003.

\bibitem{minguzzi2015raychaudhuri}
E.~Minguzzi, {\it Raychaudhuri equation and singularity theorems in finsler
  spacetimes},  {\em Classical and Quantum Gravity} {\bf 32} (2015), no.~18
  185008.

\bibitem{kunzinger2015hawking}
M.~Kunzinger, R.~Steinbauer, M.~Stojkovi{\'c}, and J.~A. Vickers, {\it
  Hawking's singularity theorem for c1, 1-metrics},  {\em Classical and Quantum
  Gravity} {\bf 32} (2015), no.~7 075012.

\bibitem{kunzinger2015penrose}
M.~Kunzinger, R.~Steinbauer, and J.~A. Vickers, {\it The penrose singularity
  theorem in regularity c1, 1},  {\em Classical and Quantum Gravity} {\bf 32}
  (2015), no.~15 155010.

\bibitem{sarma2016axially}
D.~Sarma, F.~Ahmed, and M.~Patgiri, {\it Axially symmetric, asymptotically flat
  vacuum metric with a naked singularity and closed timelike curves},  {\em
  Advances in High Energy Physics} {\bf 2016} (2016).

\bibitem{ahmed2017cylindrically}
F.~Ahmed, {\it Cylindrically symmetric, asymptotically flat, petrov type d
  spacetime with a naked curvature singularity and matter collapse},  {\em
  Advances in High Energy Physics} {\bf 2017} (2017).

\bibitem{olmo2015geodesic}
G.~J. Olmo, D.~Rubiera-Garcia, and A.~Sanchez-Puente, {\it Geodesic
  completeness in a wormhole spacetime with horizons},  {\em Physical Review D}
  {\bf 92} (2015), no.~4 044047.

\bibitem{mkenyeleye2014gravitational}
M.~D. Mkenyeleye, R.~Goswami, and S.~D. Maharaj, {\it Gravitational collapse of
  generalized vaidya spacetime},  {\em Physical Review D} {\bf 90} (2014),
  no.~6 064034.

\bibitem{stoica2012schwarzschild}
O.~C. Stoica, {\it Schwarzschild's singularity is semi-regularizable},  {\em
  The European Physical Journal Plus} {\bf 127} (2012), no.~7 1--8.

\bibitem{liu2014singularity}
Y.~Liu, D.~Malafarina, L.~Modesto, and C.~Bambi, {\it Singularity avoidance in
  quantum-inspired inhomogeneous dust collapse},  {\em Physical Review D} {\bf
  90} (2014), no.~4 044040.

\end{thebibliography}

\end{document}